\documentclass{article}
\DeclareMathAlphabet\mathbfcal{OMS}{cmsy}{b}{n}
\usepackage{spconf,amsmath,graphicx,bm,amssymb}

\title{SIMD-SIZE AWARE WEIGHT REGULARIZATION \\ FOR FAST NEURAL VOCODING ON CPU}
\name{Hiroki Kanagawa and Yusuke Ijima}
\address{NTT Corporation}

\begin{document}
%
\maketitle
\begin{abstract}
This paper proposes weight regularization for a faster neural vocoder.
Pruning time-consuming DNN modules is a promising way to realize a real-time vocoder on a CPU ({\it e.g.} WaveRNN, LPCNet).
Regularization that encourages sparsity is also effective in avoiding the quality degradation created by pruning.
However, the orders of weight matrices must be contiguous in SIMD size for fast vocoding.
To ensure this order, we propose explicit SIMD size aware regularization.
Our proposed method reshapes a weight matrix into a tensor so that the weights are aligned by group size in advance, and then computes the group Lasso-like regularization loss.
Experiments on 70\% sparse subband WaveRNN show that pruning in conventional Lasso and column-wise group Lasso degrades the synthetic speech's naturalness.
The vocoder with proposed regularization 1) achieves comparable naturalness to that without pruning and 2) performs meaningfully faster than other conventional vocoders using regularization.
\end{abstract}
\section{Introduction}
\label{sec:introduction}
The neural vocoder represented by WaveNet \cite{aaron_wavenet_corr2016} has dramatically improved the quality of text-to-speech (TTS) synthesis.
While WaveNet can generate high-quality speech waveforms directly from conditioning features via a large causal convolution-based autoregressive model, its huge computational cost and autoregressive (AR) architecture prevent fast vocoding.
To allow easier parallel computation, many neural vocoders based on non-autoregressive structures have been proposed \cite{VanDenOord2018,Ping2019,RyanPrengerRafaelValle2019,yamamoto2020parallel,Kumar2019melgan,Kong2020hifigan}.
These schemes offer high processing speeds if the device is specialized for parallel computing, such as GPUs.
On the other hand, to achieve fast neural vocoding on CPUs, the computational complexity must be reduced drastically.
WaveRNN achieves real-time vocoding by replacing huge causal convolutions of WaveNet with a simple GRU, and pruning its weights \cite{Kalchbrenner2018}.
LPCNet also introduces signal processing insights into the speech generation process and reduces the number of DNN parameters from that of WaveRNN \cite{Valin2018}.
Other approaches to reduce the number of DNN inferencing iterations, with multi-sample generation in a single forward propagation step \cite{Popov2020multisample,patrick2020multisample} and prediction of shortened subband signals instead of waveforms \cite{okamoto2017subbandwavenet,Yu2020}.

Quantization and low-rank approximation are promising alternatives to pruning for paring the DNN module's computational complexity.
\cite{Yu2020} roughly quadrupled speeds by quantizing neural vocoder weights to 8-bit integers, but it requires quantization error-aware training and an appropriate intrinsic implementation for integers, resulting in high implementation cost and hardware dependency.
Although the low-rank approximation shrinks the model size, its speed-up contribution is limited due to the increased number of matrices that must be computed, which requires more matrix-vector product instructions ({\it e.g. ``gemv''}) to be called \cite{kanagawa2020tensordecomp}.
Pruning \cite{hao2017prune,zhu2018prune} used in WaveRNN and LPCNet substitutes the elements of the weight matrix with zeros in the training process.
During inference, the calculation of zero weights can be skipped, and hardware dependency is also small because the weights can be treated as floating-point without modification.
While excessive pruning leads to quality degradation, regularization is an effective solution.
Lasso regularization promotes a sparse model, thus allowing for a smaller gap in the model with and without pruning \cite{thom2017lassoreg}.
However, the order of non-zero weight elements becomes non-contiguous when using Lasso for regularization.
To exploit fully the fast single instruction multiple data (SIMD) operations intrinsic to CPUs, the non-zero elements must be contiguous, so Lasso is not optimal for SIMD.
The use of group Lasso (gLasso) \cite{scardapane2017grouplassoreg} allows for a sequence of non-zero elements, but the model's expressiveness is sacrificed due to row- or column-wise weight sparsity.

To ensure both speed and quality, we propose a gLasso-like regularization approach that explicitly sparsifies weights with the group pruning size.
The proposed method aligns the weights in both the non-zero and zero regions with enough size to fully occupy the SIMD registers at once.
Thus, group pruning across the boundary between regions can be avoided, and computational efficiency can be improved while maintaining model expressiveness.
In our experiments, we used multi-sample subband WaveRNN \cite{kanagawa2022multisample} with 70\% sparsity.
While pruning degraded the naturalness of the synthetic speech of the vocoder without regularization and the one with the use of Lasso and gLasso regularization, regularization with the proposed method and pruning maintained the same naturalness as that achieved before pruning. 
We also found that our vocoder was faster than the conventional alternatives.

\section{Multi-sample subband WaveRNN}
\label{sec:format}
\subsection{Model architecture and its training}
\begin{figure}[t]
\begin{center}
\includegraphics[width=1.0\columnwidth]{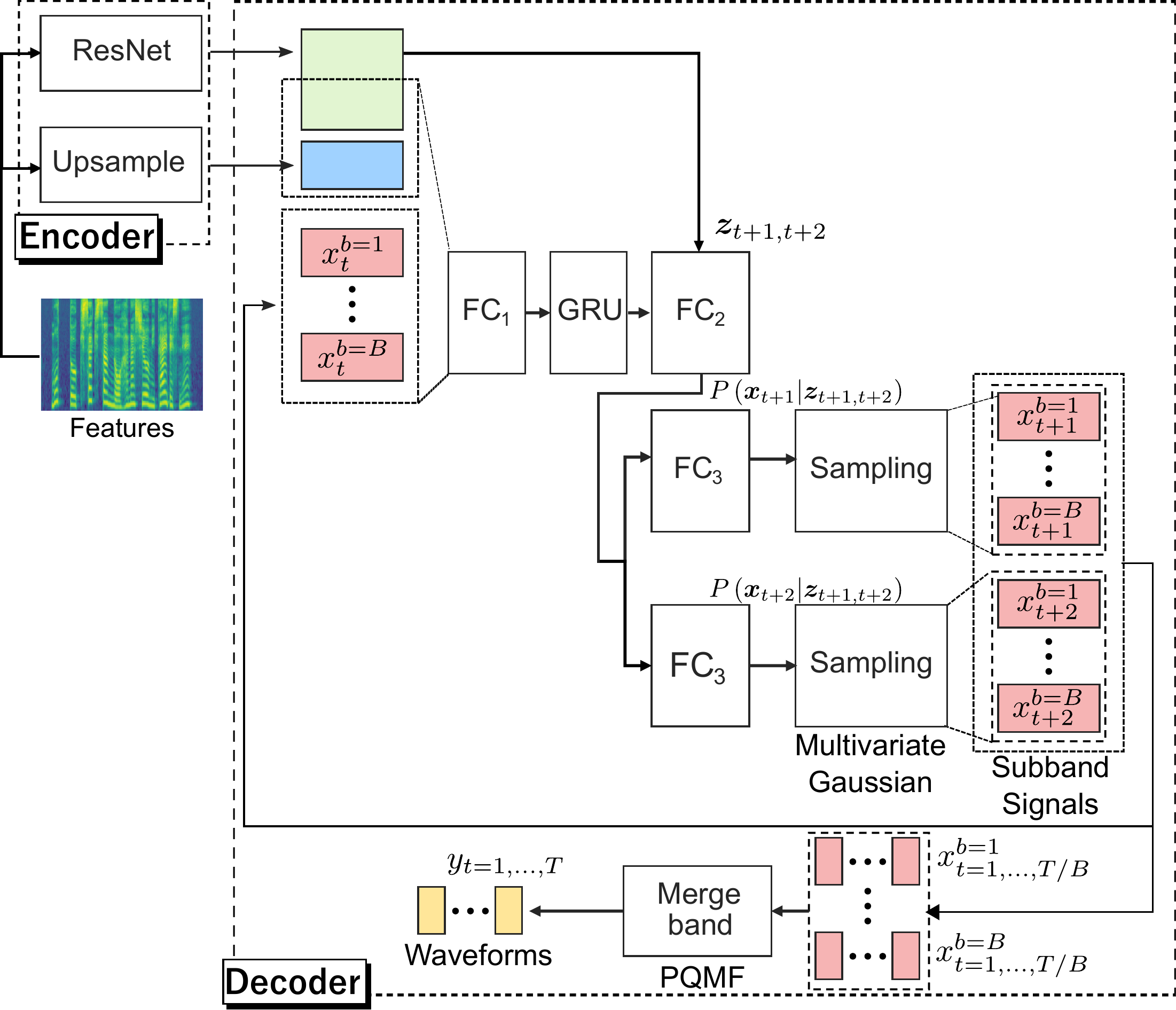}
\end{center}
\vspace{-5mm}
\caption{Overview of multi-sample subband WaveRNN via multivariate Gaussian \cite{kanagawa2022multisample}. This vocoder predicts $M$ subband signals simultaneously in single forward propagation step ($M$=2 is used in this figure).}
\label{fig:conventional_wavernn}
\end{figure}
Subband WaveRNN \cite{Yu2020} reduces sequence length from $T$ to $T/B$ by predicting $B$-band subband signals instead of a speech waveform.
\cite{kanagawa2022multisample} extended it to multi-sample generation for even faster vocoding; Fig.~\ref{fig:conventional_wavernn} shows an overview.
The model consists of an encoder and a decoder, which are responsible for frame rate and sample rate, respectively.
Encoders upsample frame-level acoustic features to corresponding samples.
The decoder generates predictions of the next time $t+\tau \forall _m \in [1,M - 1]$ from the output of the encoder and the previous $M$ subband signals, where $m$ is the index of the number of subband signals to be generated simultaneously.
To generate multiple samples in single forward propagation, linear layer FC3s are provided for subband signals of 
$t+\tau \forall _m  \in [1,M - 1]$.
This module jointly predicts the associations among subband signals because PQMF has band overlaps.
Assuming a multivariate Gaussian as FC3's target, this vocoder minimizes the negative log-likelihood given by:

\footnotesize
\vspace{-3mm}
\begin{align}
{\cal L}_{{\rm NLL}}\left( \theta  \right) =  - \sum\limits_{t = 1}^{T/B} {\sum\limits_{m = 1}^M {\ln {\cal N}\left( {{\bm x}_{t + m} ;{\bm \mu }\left( {{\bm z}_{t + \tau \forall _m } ,\theta } \right),{\bm \Sigma }\left( {{\bm z}_{t + \tau \forall _m } ,\theta } \right)} \right)} } ,
\label{eq:nll_mvgauss}
\end{align}
\vspace{-1mm}
\normalsize
where $\theta$, ${\bm z}_t$, ${\bm x}_t\in {\mathbb R}^B$, ${\bm \mu } \in {\mathbb R}^B$ and ${\bm \Sigma } \in {\mathbb R}^{B \times B}$ are the DNN model parameters, FC2's output, subband signals, the mean vector and covariance matrix of the multivariate Gaussian, respectively.
To guarantee spectral reproducibility, STFT loss ${\cal L}_{{\rm STFT}} \left( \theta  \right)$
\cite{takaki2019stft_loss} is calculated by generating subband signals from the multivariate Gaussian via a reparameterization trick.
This is added to Eq.~\eqref{eq:nll_mvgauss} without scaling to optimize the vocoder.

\subsection{Weight pruning}
For fast vocoding, pruning is performed during training.
We performed pruning by gradually increasing sparsity \cite{zhu2018prune} in the same manner as WaveRNN using:

\small
\vspace{-3mm}
\begin{align}
d_s  = d\left[ {1 - \left\{ {1 - \left( {s - s_0 } \right)/S} \right\}^3 } \right],
\label{eq:prune}
\end{align}
\vspace{-1mm}
\normalsize
where $s$, $s_0$, and $S$ are the current-, start-, and total- pruning step, respectively.
$d$ is the target density, thus the sparsity is defined as $1 - d$.
To fully utilize vector algebra with SIMD, we apply group pruning \cite{hao2017prune}.
The group size of FC1, GRU, and FC2 for pruning was set to 16.
Taking Intel's AVX2 intrinsic instruction set as SIMD, the dot product can be calculated for 16 elements in two SIMD operations.
This is done by putting eight 32-bit float elements on a register and calculating the dot product, which is then applied to and added to the other eight neighboring elements.

\subsection{Weight regularization}
\label{sec:regularization}
In order to avoid a degradation in model expressiveness due to pruning, regularization is an efficient approach to sparsify the model in advance.
A well-known Lasso regularization term is computed as follows:
\small
\vspace{-3mm}
\begin{align}
{\cal L}_{{\rm Reg}}^{\textsc{Lasso}} \left( \theta  \right) = \sum\limits_r {\sum\limits_{i = 1}^{I_r } {\sum\limits_{j = 1}^{J_r } {\left| {W_r \left( {i,j} \right)} \right|} } } ,
\end{align}
\vspace{-1mm}
\normalsize
where $r$, ${\bm W}_r \in {\mathbb R}^{I_r  \times J_r }$ are the DNN module's index and the weight matrix to be regularized, respectively.
$i$ and $j$ denote the row and column indices.

The regularization term of column-wise group Lasso is given by:
\small
\begin{align}
{\cal L}_{{\rm Reg}}^{\textsc{gLasso}} \left( \theta  \right) = \sum\limits_r {\sum\limits_{j = 1}^{J_r } {\left\| {{\bm w}_r^i }, \right\|} }
\end{align}
\normalsize
where ${{\bm w}_r^i }$ and $\left\|  \cdot  \right\|$ are the ${\bm W}_r$'s $i$-th column vector and the L2 norm operator, respectively.
The final objective function with regularization is reformulated as:

\small
\begin{align}
{\cal L}\left( \theta  \right) = {\cal L}_{{\rm NLL}} \left( \theta  \right) + {\cal L}_{{\rm STFT}} \left( \theta  \right) + \lambda {\cal L}_{{\rm Reg}} \left( \theta  \right),
\label{seq:final_objective}
\end{align}
\normalsize
where $\lambda$ is the scale for the regularization term ${\cal L}_{{\rm Reg}} \left( \theta  \right)$.

\section{Proposed SIMD-size aware weight reguralizaton}
\label{sec:proposed}
The Lasso and gLasso regularizations described in Section \ref{sec:regularization} could suffer from group pruning of sparse weights across the zero/non-zero boundary, resulting in quality degradation.
To overcome this problem, we propose a regularization that assumes group pruning at the SIMD size.
The weight matrix ${\bm W}_r$ is reshaped in advance into a third-order tensor ${\mathbfcal W}_r  \in {\mathbb R}^{I_r  \times J_r /G \times G}$ to make the sparse weight's block size equal to group pruning size $G$.
Then, the proposed regularization term for this tensor is formulated as follows:

\small
\begin{align}
{\cal L}_{{\rm Reg}}^{\textsc{Proposed}} \left( \theta  \right) = \sum\limits_r {\sum\limits_{i = 1}^{I_r } {\sum\limits_{j = 1}^{J_r } {\sqrt {\sum\limits_{g = 1}^G {{\cal W}_r \left( {i,j,g} \right)^2 } } } } } .
\label{seq:proposed_reg}
\end{align}
\normalsize

\begin{figure}[t]
\begin{center}
\includegraphics[width=0.85\columnwidth]{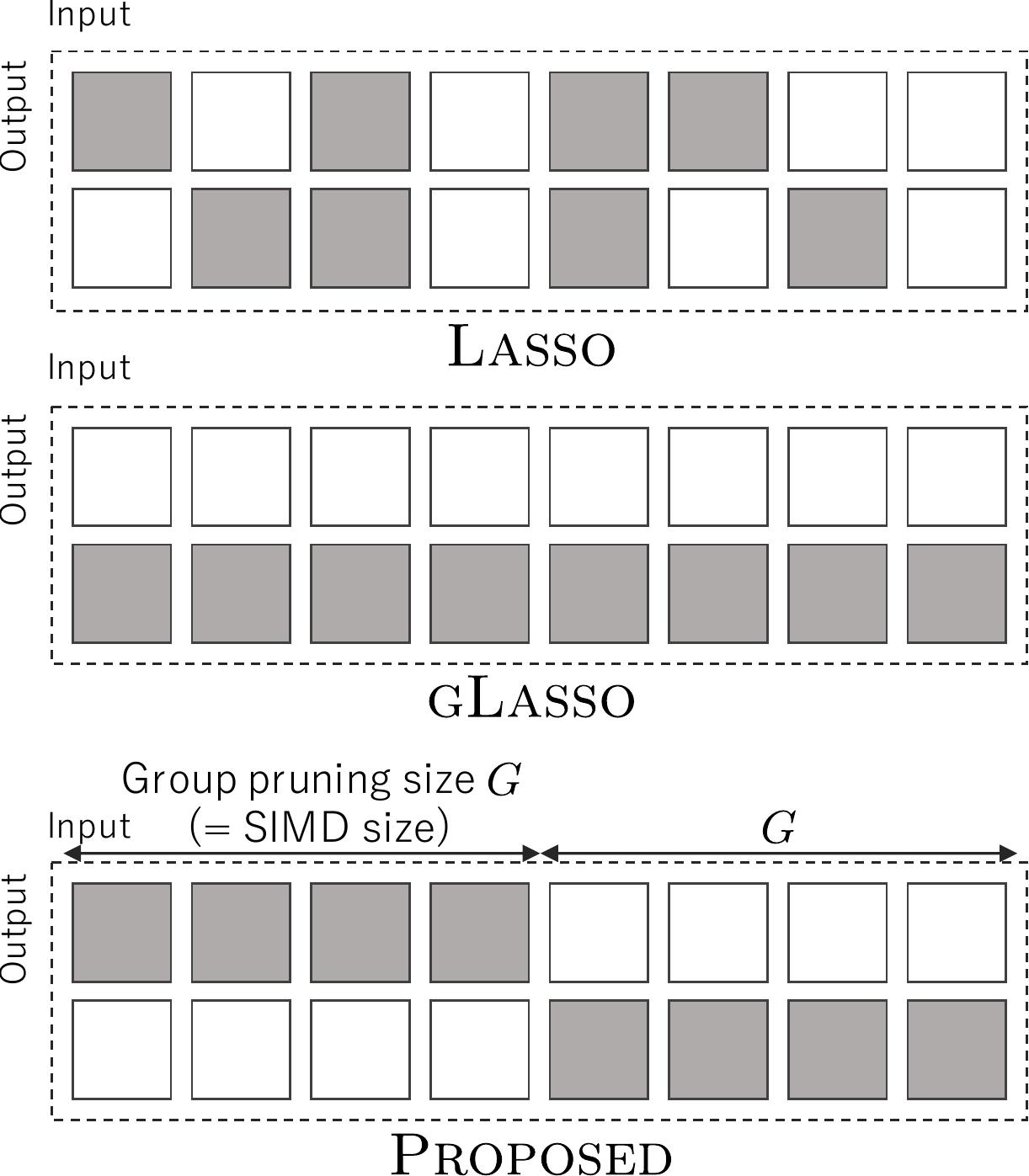}
\end{center}
\normalsize
\vspace{-6mm}
\caption{Comparison between Lasso, gLasso, and our regularization proposal applied to a single weight matrix. The white and gray regions denote non-zero and zero components, respectively.}
\label{fig:concept}
\vspace{-2mm}
\end{figure}
Figure \ref{fig:concept} compares regularized weights when Lasso, gLasso, and our proposed method (\textsc{Proposed}) are applied.
The white and gray regions of the matrices in this figure are the non-zero and zero components, respectively.
This prior matrix-to-tensor conversion and computing Eq.~\eqref{seq:proposed_reg} contribute to SIMD size group-wise sparsification.
Therefore, group-wise sparsified weights can be preferentially pruned, reducing the probability of pruning portions contain non-zero elements.
Since weights are continuous in SIMD size, weights can always be assigned to aligned memory, which is advantageous in terms of computing efficiency.

\section{Experiments}
\label{sec:expriments}

\subsection{Setup}
We used speech data uttered by a Japanese professional female speaker. The sampling frequency was 22.05 kHz.
200 utterances were extracted as evaluation data (18.3 minutes), and the remainder were used for training and validation (30.6 hours).

Eighty-dimensional logarithmic mel-spectrograms were used as the conditioning feature of the neural vocoder.
The analysis frame shift was 5 ms\footnote{Although we also investigated the commonly used frame shift of 12.5 ms in our preliminary experiments, we chose to set it to 5 ms because it better reproduced the pitch of synthetic speech. If a faster inference speed is preferred, the frame shift can be set to 12.5 ms like other studies for lower encoder computational complexity.}.
The ResNet of the encoder has ten residual blocks, each consisting of 1D-convolution with 128 units, batch normalization, and activation.
ReLU was used for all activations, and the simultaneous generation sample was set to two likewise \cite{kanagawa2022multisample}.
This multi-sample vocoder occasionally failed to predict accurate variance parameters which yielded clicking sounds.
To avoid this problem, we 1) eliminated variance outliers and 2) clipped sampled results in a similar way to \cite{Popov2020multisample}.
The number of training steps was 5000k, and parameters for pruning steps in Eq.~\eqref{eq:prune} were set to $S_0=2000$k and $S=2500$k.
We used $d=0.3$ and $\lambda$ in Eq.~\eqref{seq:final_objective} to $1.0\times10^{-4}$ for all regularization methods.
This guarantees that the model sparsity is at least 70\%.
If the number of zero elements increases due to regularization, the sparsity could be higher than this.
The vocoder's optimization was performed using RAdam \cite{liu2020variance}, with $\alpha = 1.0 \times 10^{ - 4}$, $\beta = \left( {0.9,{\kern 1pt} 0.999} \right)$, and $\varepsilon  = 1.0 \times 10^{-8}$.

\subsection{Weight heatmap comparison}
\label{sec:heatmap_comparison}
\begin{figure}[t]
\begin{center}
\includegraphics[width=0.92\columnwidth]{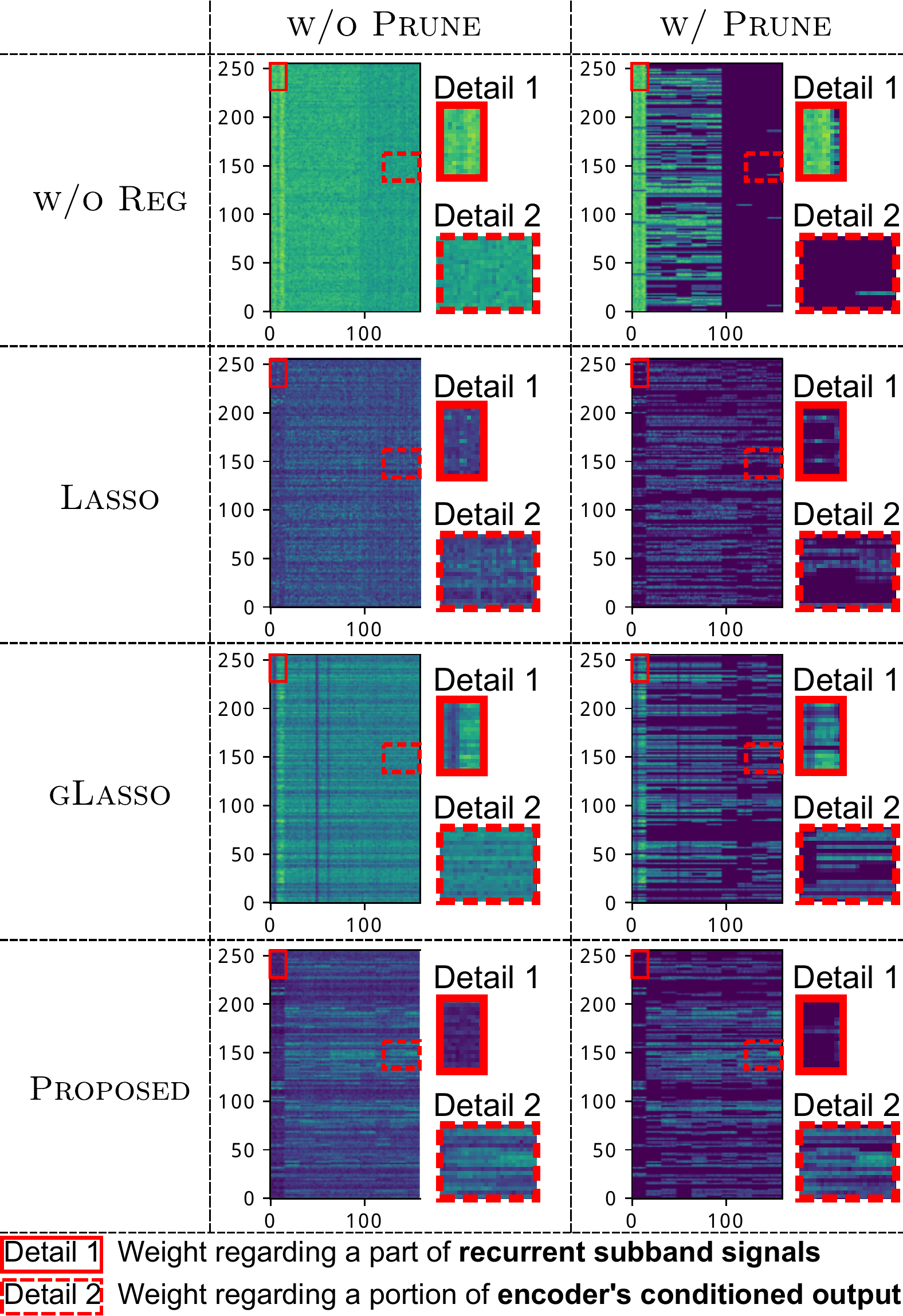}
\end{center}
\normalsize
\vspace{-1mm}
\caption{FC1's weight heatmap comparison. From top to bottom: \textsc{w/o Reg}, \textsc{Lasso}, \textsc{gLasso}, and \textsc{Proposed}. The right and left heatmaps show without and with pruning, respectively. The left-side of the heatmaps receive the recurrent multiple subband signals (\textit{e.g.} ``Detail 1'').
Other right-side of ones accept the encoder's conditioned output (\textit{e.g.} ``Detail 2'').
The comparison of ``Detail 1'' and ``2'' between \textsc{w/o} and \textsc{w/ Prune}, demonstrated that \textsc{Proposed} kept the minimum difference among them.}
\label{fig:heatmap_comparison}
\end{figure}
Figure \ref{fig:heatmap_comparison} shows FC1 (described in Fig.~\ref{fig:conventional_wavernn}) weight heatmaps before and after pruning for each method.
The horizontal and vertical axes are the input and output dimensions, respectively.
Comparing \textsc{w/o Reg (w/o Prune)} and \textsc{w/o Reg (w/ Prune)}, we can see that the elements have been replaced with zero leaving large components.
As described in Fig.~\ref{fig:conventional_wavernn}, the recurrent multiple subband signals are fed to FC1, so these weights are particularly large on the left-side of the heatmap as if to focus on their signals (\textit{e.g.} ``Detail 1'' in Fig.~\ref{fig:heatmap_comparison}).
Other right-side weight components are responsible for receiving encoder's ouputs (\textit{e.g.} ``Detail 2'').
These results revealed that \textsc{w/o Reg (w/ Prune)} disregarded or neglected the encoder’s conditioned outputs.
Although \textsc{Lasso (w/o Prune)} was not overly dependent on the most recent sample, its weights had discontinuous order.
The \textsc{gLasso (w/o Prune)} yield poor sparsification, because under the constraint of sparsifying entire columns, it was difficult to find compact representations.
These regularizations, like \textsc{w/o Reg (w/o Prune)}, are prone to degrade the synthesized speech's quality because group pruning is applied across the boundaries between zero and non-zero.
On the other hand, the proposed method \textsc{Proposed (w/o Prune)} was sparse and its weights were continuously aligned in group pruning size $G=16$ in the column direction before pruning.
Since there is no drastic difference in the heatmaps between with and without pruning, we found that pruning can be done without sacrificing the model's expressiveness.

\subsection{Subjective evaluations}
We subjectively evaluated the naturalness of synthetic speech by using mean opinion score (MOS) on a five-point scale ranging from 5: very natural to 1: very unnatural. Sixty listeners participated in the test via crowdsourcing.
They evaluated ten sentences for each method, randomly selected from all 200 evaluation data, for a total of sixty sentences.
These participants were different evaluators for analytic resynthesis and TTS.

\subsubsection{Vocoding for extracted acoustic features from natural speech}
\begin{figure}[t]
\begin{center}
\includegraphics[width=0.9\columnwidth]{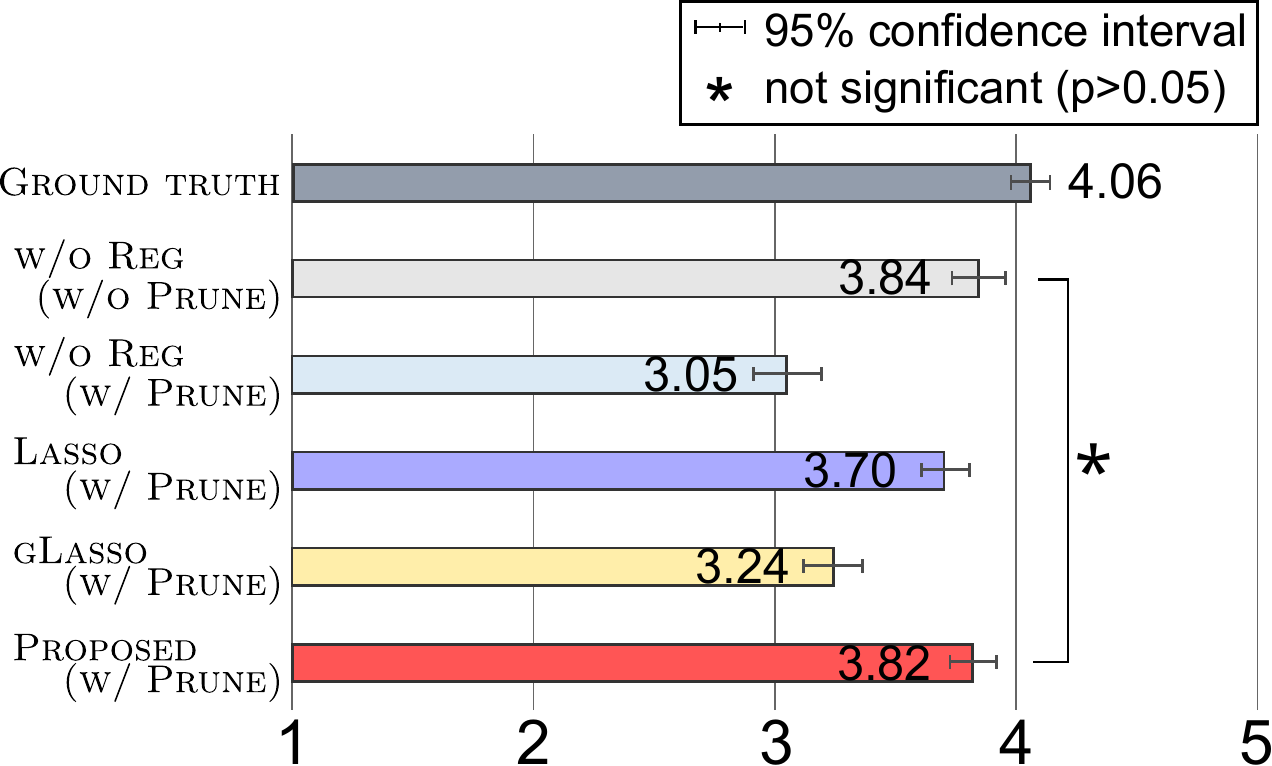}
\end{center}
\vspace{-7mm}
\caption{Mean opinion scores of naturalness.
Acoustic features for vocoding were extracted from natural speech.}
\vspace{-2mm}
\label{fig:eva_mos_resyn}
\end{figure}
\begin{figure}[t]
\begin{center}
\includegraphics[width=0.9\columnwidth]{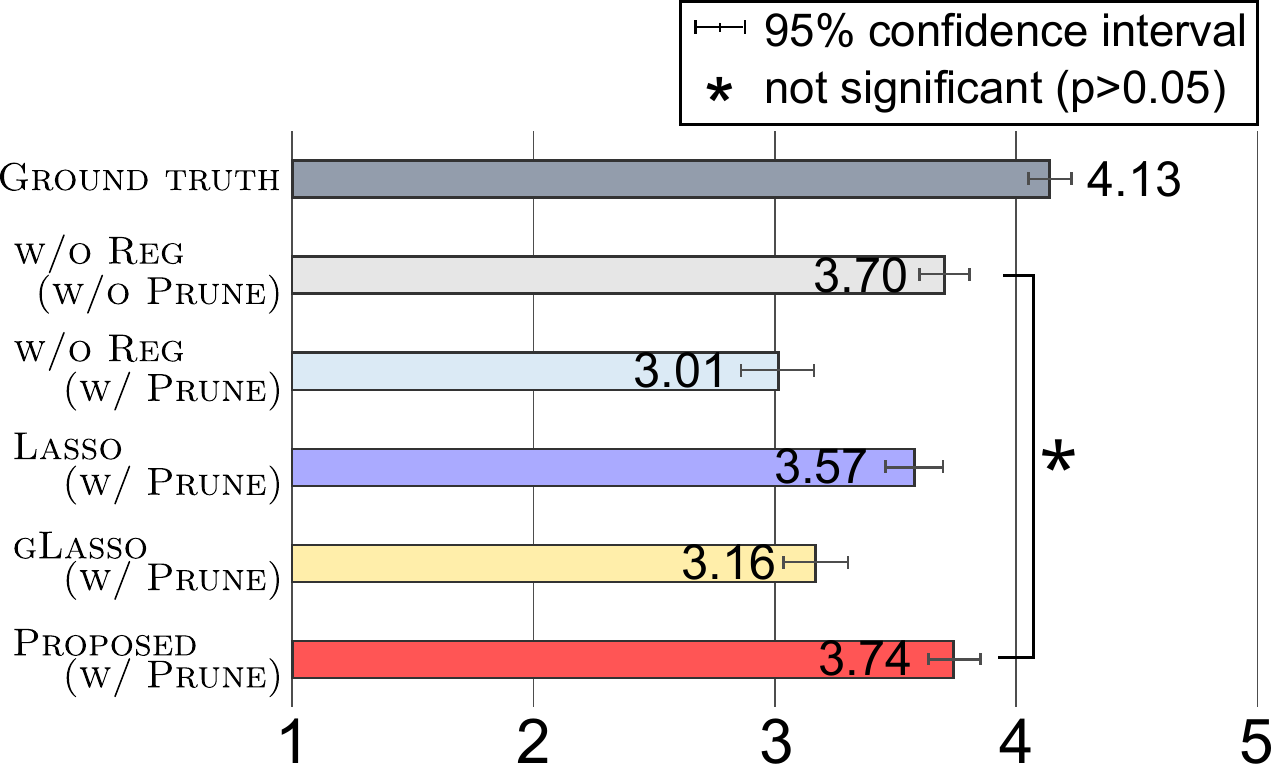}
\end{center}
\vspace{-7mm}
\caption{Mean opinion scores in terms of naturalness.
Acoustic features for vocoding were predicted by the TTS model.}
\label{fig:eva_mos_tts}
\end{figure}

Figure \ref{fig:eva_mos_resyn} shows the subjective evaluation results of vocoding with acoustic features extracted from natural speech.
\textsc{w/o Reg (w/o Prune)} yielded lower performance than Ground truth, but still obtained high naturalness.
\textsc{w/o Reg (w/ Prune)} degraded significantly more than \textsc{w/o Reg (w/o Prune)}.
This is due to 1) neglecting the conditioning spectral information in pruning, and 2) excessive reliance on recurrent samples led to quality degradation due to mismatches with the training time; as described in Section \ref{sec:heatmap_comparison}.
The performance of \textsc{gLasso (w/ Prune)} also falls for the same reason, just not as much as \textsc{w/o Reg (w/ Prune)}.
On the other hand, \textsc{Lasso (w/ Prune)} showed no more significant degradation from \textsc{w/o Reg}, even with pruning.
This is due to the fact that model does not rely excessively on recent samples, but focuses more on the conditioning spectral information.
\textsc{Proposed (w/ Prune)} outperforms these pruned models and achieves naturalness comparable to that of \textsc{w/o Reg (w/o Prune)}.

\subsubsection{Vocoding for acoustic features predicted by TTS}
To investigate robustness against degraded acoustic features, FastSpeech2 \cite{Ren2021fastspeech2} as the TTS model was also trained with the same data as the neural vocoders.
We fed 380 kinds of symbols including phoneme and prosodic information to FastSpeech2.
It was optimized via a minimum mean absolute error criterion with 2000k steps by Adam \cite{kingma_adam_iclr_2015} with $\beta = \left( {0.9,{\kern 1pt} 0.98} \right)$, and $\varepsilon  = 1.0 \times 10^{-9}$.
We followed the same learning rate schedule in \cite{vaswani2017transformer}.

Figure \ref{fig:eva_mos_tts} shows the subjective evaluation results.
The overall difference in scores between ground truth and synthetic speech was greater when using features extracted from natural speech since acoustic features predicted by TTS were degraded from the original one.
The \textsc{Proposed (w/ Prune)} matched the naturalness of \textsc{w/o Reg (w/o Prune)}.
This result confirms that the regularization proposal is robust to acoustic features degraded by TTS.

\subsection{Vocoding speed comparison}
The real-time factors (RTFs) were calculated to measure the inference speeds for all methods.
RTF is defined by:
\vspace{-3mm}

\small
\begin{align}
{\rm RTF} := T_{{\rm inference}} /T_{{\rm data}},
\end{align}
\normalsize
where $T_{{\rm data}} $ and $T_{{\rm inference}} $ are speech length and single-thread inference time measured on an Intel Core i7-8750H CPU 2.20 GHz, respectively.

\begin{figure}[!t]
\begin{center}
\includegraphics[width=0.9\columnwidth]{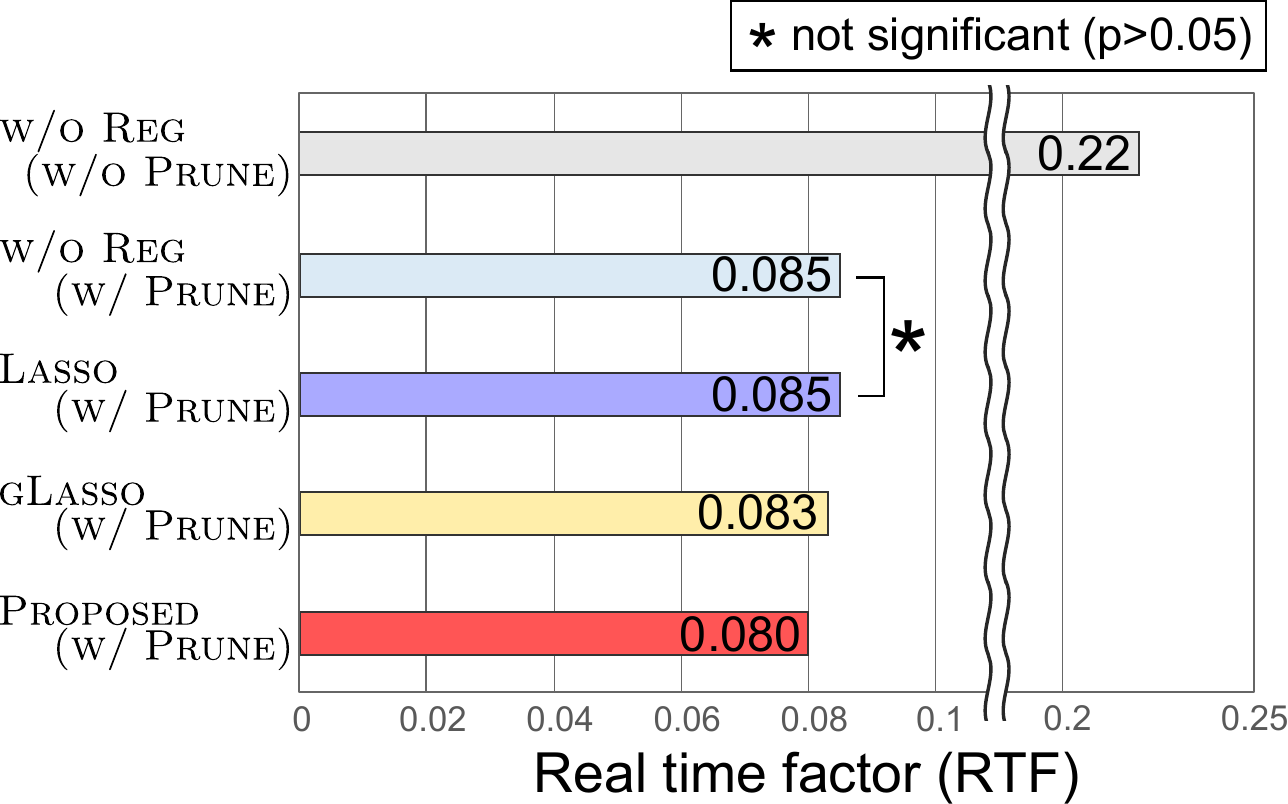}
\end{center}
\vspace{-6mm}
\caption{Average RTFs obtained from all evaluation data.}
\label{fig:speed_comparison}
\end{figure}
Figure~\ref{fig:speed_comparison} shows averaged RTFs from all evaluation data.
\textsc{w/o Reg (w/ Prune)} yielded a significant speed improvement over \textsc{w/o Reg (w/o Prune)} owing to pruning.
\textsc{Lasso (w/ Prune)} showed no speed improvement from \textsc{w/o Reg (w/ Prune)}.
As discussed in Section \ref{sec:heatmap_comparison}, this was attributed to Lasso regularization promoting non-contiguous sparse matrices, which fails to yield a SIMD-friendly contiguous sparse matrix.
On the other hand, \textsc{gLasso (w/ Prune)} slightly improved the RTF.
The reason is the 70\% sparsity by pruning, further increased column-wise contiguous zeroed regions via regularization.
The proposed method achieved better RTFs than \textsc{gLasso (w/ Prune)}, since it achieved a group-wise sparser matrix than \textsc{gLasso}, as mentioned in Section \ref{sec:heatmap_comparison}.
Our RTF=0.080 is nearly comparable to the one of the recent non-AR HiFi-GAN (v3) \cite{Kong2020hifigan}, which works fast on CPUs (RTF=0.075).
Since they used a higher clock CPU (Intel Core i7 CPU 2.6 GHz) than ours, our proposed method might outperform their speed if on the same CPU.
Furthermore, the combination of the other faster approach for WaveRNN ({\textit e.g.} \cite{kanagawa2022multisamplejoint}) and the proposed regularization would be able to provide a significant speed-up compared to HiFi-GAN.

\section{Conclusion}
\label{sec:conclusion}
In this work, we proposed SIMD-size aware group-wise regularization to avoid the quality degradation associated with neural vocoder pruning.
We incorporated the regularization proposal into a subband WaveRNN-based vocoder and showed that the regularized weights have group-wise continuous orders suitable for SIMD computation.
No major differences in our vocoder's weight layout were observed via heatmaps of before and after pruning.
Subjective evaluations regarding naturalness demonstrated that the proposed pruned vocoder outperformed that with no regularization, Lasso, and group Lasso.
In particular, our vocoder achieved comparable naturalness to that achieved without pruning.
A speed evaluation also revealed that our vocoder performed significantly faster than the existing alternatives.

Our method can also increase contiguous zeroed region more efficiently than \textsc{Lasso} and \textsc{gLasso} even if regularized to fit processors with smaller SIMD sizes, \textit{e.g.} Intel SSE4 and Arm NEON.
So we expect to run significantly faster than them without any loss of quality.
Our first future work will apply the proposed method for Arm NEON and confirm its efficiency on embedded processors.
Our second second future work will compare and combine our regularization with AlignReg \cite{neill2022alignreg}, which has similar concepts proposed for natural language processing.
Since the proposed regularization is not limited to neural vocoders, we will also plan to apply it to other computationally expensive models (\textit{e.g.} RNN-T \cite{graves2012rnnt} and BERT \cite{devlin2018bert}).

\newpage
\bibliographystyle{IEEEbib}
\bibliography{mybib}

\end{document}